\documentclass[sigconf,screen]{acmart}
\AtBeginDocument{%
  }

\setcopyright{acmlicensed}
\copyrightyear{2026}
\acmYear{2026}
\setcopyright{cc}
\setcctype{by}
\acmConference[SESAME '26]{4th Workshop on Serverless Systems, Applications and Methodologies }{April 27--30, 2026}{Edinburgh, Scotland Uk}
\acmBooktitle{4th Workshop on Serverless Systems, Applications and Methodologies (SESAME '26), April 27--30, 2026, Edinburgh, Scotland Uk}
\acmDOI{10.1145/3802803.3802911}
\acmISBN{/2026/04}
\usepackage{microtype}
\usepackage{graphicx}
\usepackage{subfigure}
\usepackage{booktabs} % for professional tables
\usepackage{mathtools} 
\usepackage{amsmath}
\usepackage{siunitx}
\usepackage{enumitem}

\usepackage{xurl}

\usepackage{colortbl}

\usepackage{makecell}
\usepackage{multicol}
\usepackage{pifont}
\definecolor{lightergray}{RGB}{230,230,230}
\definecolor{DarkGreen}{RGB}{30,130,30}

\usepackage[ruled,vlined]{algorithm2e}

\begin{document}

\title{OmniInfer: System-Wide Acceleration Techniques for Optimizing LLM Serving Throughput and Latency}

% \affiliation{%

\author{Jun Wang, Yunxiang Yao, Wenwei Kuang, Runze Mao,
Zhenhao Sun, Zhuang Tao, Xingyuan Chen, \\ Ziyang Zhang,
Dengyu Li, Jiajun Chen, Yu Gao, Changjian Zhang,
Chengda Wu, Meng Wang,\\ Yishan Wu, Zhili Wang,
Kai Cui, Congzhi Cai, Weixi Zhang, Longwen Lan,
Wei Han, Ken Zhang}

\affiliation{%
  \institution{Theory Lab - Leibniz, Central Research Institute, 2012 Labs, Huawei Technology Co., Ltd.}
  \city{Hong Kong}
  \country{China}
}

\email{{wang.jun7, ken.zhang1}@huawei.com}

\renewcommand{\shortauthors}{Wang et al.}

\begin{abstract}
Large Language Models drive a wide range of modern AI applications but impose substantial challenges on large-scale serving systems due to intensive computation, strict latency constraints, and throughput bottlenecks.
We introduce {OmniInfer}, a unified system-level acceleration framework designed to maximize end-to-end serving efficiency through fine-grained optimization of expert placement, cache compression, and scheduling.
OmniInfer integrates three complementary components: {OmniPlacement} for load-aware Mixture-of-Experts scheduling, {OmniAttn} for sparse attention acceleration, and {OmniProxy} for disaggregation-aware request scheduling.
Built atop vLLM, OmniInfer delivers system-wide performance gains through adaptive resource disaggregation, efficient sparsity exploitation, and global coordination across prefill and decode phases.
Evaluated on DeepSeek-R1 within a 10-node Ascend 910C cluster, OmniInfer achieves a 52\% throughput gain, reaching 616 QPM.
Within this unified framework, TPOT is reduced by up to 36\%, and the superimposed OmniProxy further reduces TTFT by up to 38\%.
% The project is open-sourced at \url{https://gitee.com/omniai/omniinfer}.
\end{abstract}





\maketitle

\section{Introduction}
\label{sec:introduction}

Large Language Models (LLMs) like DeepSeek~\cite{guo2025deepseek}, Qwen~\cite{yang2025qwen3}, Kimi~\cite{team2025kimi} and GPT~\cite{agarwal2025gpt}  have rapidly become the foundation of modern AI applications, powering chatbots~\cite{achiam2023gpt}, tool use~\cite{schick2023toolformer}, search~\cite{zilliztech2025deepsearcher}, and multimodal assistants~\cite{google2025gemini_image_editing}. 
However, their deployment at scale remains challenging due to enormous computational demands and stringent service-level objectives (SLOs), such as low time-to-first-token (TTFT) and high query-per-minute (QPM) throughput. 
To sustain practical serving, systems must not only maximize hardware utilization but also optimize scheduling across diverse and dynamic workloads.

Existing LLM serving systems have made substantial progress in improving inference efficiency, leveraging techniques such as operator fusion~\cite{zuo2025serving}, continuous batching~\cite{kwon2023efficient}, and KV-cache reuse~\cite{zheng2024sglang}. 
However, most of these systems assume homogeneous clusters and monolithic execution, where the prefill and decode stages are tightly coupled. 
This design overlooks the distinct computational and memory characteristics of the two phases: the prefill stage is compute-intensive, dominated by large matrix multiplications, whereas the decode stage is memory- and communication-bound, constrained by KV-cache accesses and sequential token generation~\cite{patel2024splitwise,zhong2024distserve}. 
Recent variants such as chunked prefill~\cite{agrawal2024taming} attempt to reduce TTFT under continuous batching, but fundamentally shift rather than eliminate prefill–decode interference, and thus cannot fully resolve the inherent contention between the two stages.
Treating the two phases uniformly not only results in inefficient hardware utilization but also exacerbates head-of-line (HOL) blocking under dynamic workloads~\cite{zhou2024survey,pan2025survey}.

To mitigate these inefficiencies, recent work has explored {prefill–decode (PD) To mitigate these inefficiencies, recent work has explored \emph{prefill–decode (PD) disaggregation}~\cite{zhong2024distserve,patel2024splitwise}, which separates prefill and decode across heterogeneous hardware resources.
This design enables finer-grained scheduling, improved load balancing, and additional opportunities for parallelism.
However, existing PD disaggregation systems remain constrained in practice: some rely on high-speed interconnects that are not widely available, while others lack global scheduling policies that adapt effectively to dynamic workloads.
As a result, although PD disaggregation is a critical step forward, it alone is insufficient to fully address the systemic bottlenecks in modern LLM serving.

\begin{figure*}[ht]  
    \centering
    \includegraphics[width=1\linewidth]{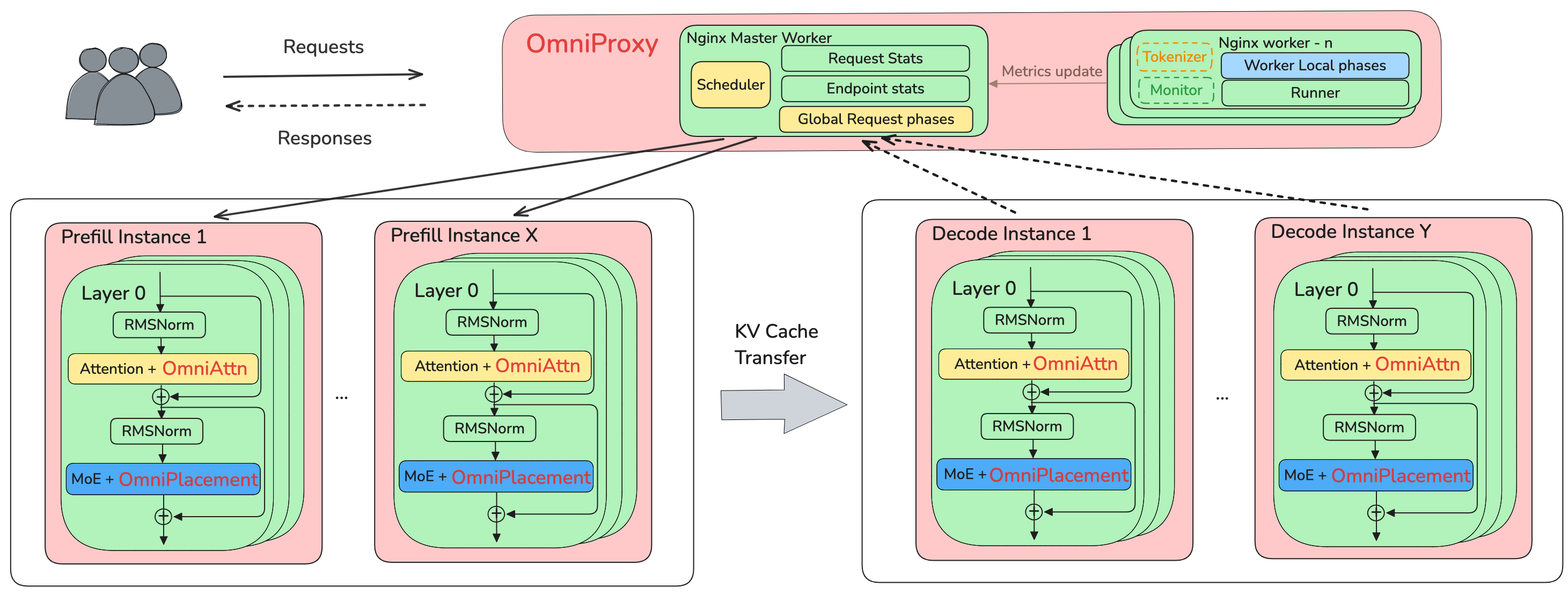}
    % \includegraphics[height=6cm,keepaspectratio]{images/omniinfer.png}
    % \vspace{-5mm}
    \caption{Structure of OmniInfer system under PD-disaggregated serving.}
    \label{fig:omniinfer}
\end{figure*}

Beyond PD disaggregation, existing serving frameworks continue to face two fundamental challenges.
First, \emph{sparse computing}, arising from mixture-of-experts (MoE) models and sparse attention mechanisms, offers theoretical efficiency gains but often introduces load imbalance and hardware under-utilization in practice~\cite{liu2024deepseek}.
Second, while several heuristic and black-box optimization~\cite{fu2024efficient,singh2025efficiently} approaches have been proposed for \emph{smart scheduling}, they generally overlook the distinct characteristics of the prefill and decode phases in LLM inference.
As LLMs scale toward larger and multimodal forms, these limitations increasingly constrain throughput, latency, and cost efficiency.

In this paper, we present {OmniInfer}, a unified suite of acceleration techniques for efficient large-scale LLM serving.
As shown in Figure~\ref{fig:omniinfer}, it consists of three lightweight components that jointly target computational and scheduling inefficiencies in disaggregated LLM serving.
This modular design decouples key serving bottlenecks into \textit{specialized} and \textit{pluggable} modules, while maintaining a unified interface that enables seamless integration with mainstream inference frameworks such as vLLM and SGLang.

Concretely, OmniInfer makes three core contributions that are modular, orthogonal and synergize with each other:
\begin{itemize}
    \item \textbf{OmniPlacement}: a load-aware MoE expert scheduling mechanism that improves efficiency in MoE inference.
    \item \textbf{OmniAttn}: optimized sparse attention that accelerates long-context inference while preserving model quality.
    \item \textbf{OmniProxy}: a scheduling proxy that coordinates prefill and decode phases using disaggregation-aware strategies to optimize throughput and latency.
\end{itemize}

We implement OmniInfer on top of a state-of-the-art serving framework vLLM~\cite{kwon2023efficient} and conduct extensive experiments with popular open-source LLMs. 
Our evaluation on a 10-node Ascend 910C cluster shows that OmniInfer achieves up to 616 QPM, where the unified framework reduces TPOT by up to 36\%, and the superimposed OmniProxy further reduces TTFT by up to 38\%.
These results establish OmniInfer as a practical, scalable, and cost-efficient solution for next-generation LLM serving.

\section{Background and Related Work}
\label{sec:related}

\paragraph{\textbf{Sparse Computing}}
The quadratic cost of self-attention in long-context LLMs has driven extensive research on sparse computation for efficient inference.
One line of work focuses on {KV-cache sparsification}, where redundant tokens are pruned to reduce memory and computation, as exemplified by H2O~\cite{NEURIPS2023_6ceefa7b} and SnapKV~\cite{NEURIPS2024_28ab4182}.
Another line modifies the {attention topology} itself: methods such as DuoAttention~\cite{xiao2024duoattentionefficientlongcontextllm}, motivated by the attention-sink phenomenon, design fixed sparse patterns that combine local windows with global anchors.
Unlike these training-dependent approaches, {OmniAttn} adopts an inference-driven perspective.
It replaces gradient-based sparsity learning with search-based compression, automatically discovering efficient attention patterns without retraining.
This decouples sparsity design from model training and enables flexible deployment across heterogeneous serving environments.

Beyond attention, Mixture-of-Experts (MoE) models introduce {expert sparsity} through token-level dynamic routing, often leading to load imbalance and hardware underutilization.
While prior systems optimize kernels~\cite{rajbhandari2022deepspeed,hwang2023tutel} or mitigate imbalance at the model level~\cite{zhou2022mixture}, they largely rely on static placement or reactive adaptation.
In contrast, {OmniPlacement} provides a proactive, closed-loop mechanism that continuously monitors expert utilization and performs lightweight, non-blocking weight migration.
This enables balanced MoE serving without architectural modification.

\paragraph{\textbf{LLM Scheduling}}
Scheduling in LLM serving is challenged by {long or shared prompts} and {heterogeneous prompt lengths}.
To exploit prompt reuse, recent systems employ KV-cache--aware prefill scheduling using radix-tree–based structures~\cite{srivatsa2025preble,hu2024memserve,zheng2024sglang}.
Meanwhile, prompt length variability induces execution bubbles that degrade both throughput and latency.
Prior work mitigates this via heuristic policies such as SJF~\cite{fu2024efficient} or by predicting request characteristics using auxiliary models~\cite{fu2024efficient,stojkovic2025dynamollm}.

Most existing schedulers address these challenges in isolation and rely either on heuristic rules~\cite{fu2024efficient,srivatsa2025preble} or black-box optimization~\cite{zhong2024distserve,singh2025efficiently}.
In contrast, {OmniProxy} jointly accounts for prompt sharing and workload heterogeneity.
By leveraging endogenous system metrics, it enables adaptive, fine-grained scheduling under dynamic serving conditions.

\begin{figure}[tb]  
    \centering
    \includegraphics[width=1\linewidth]{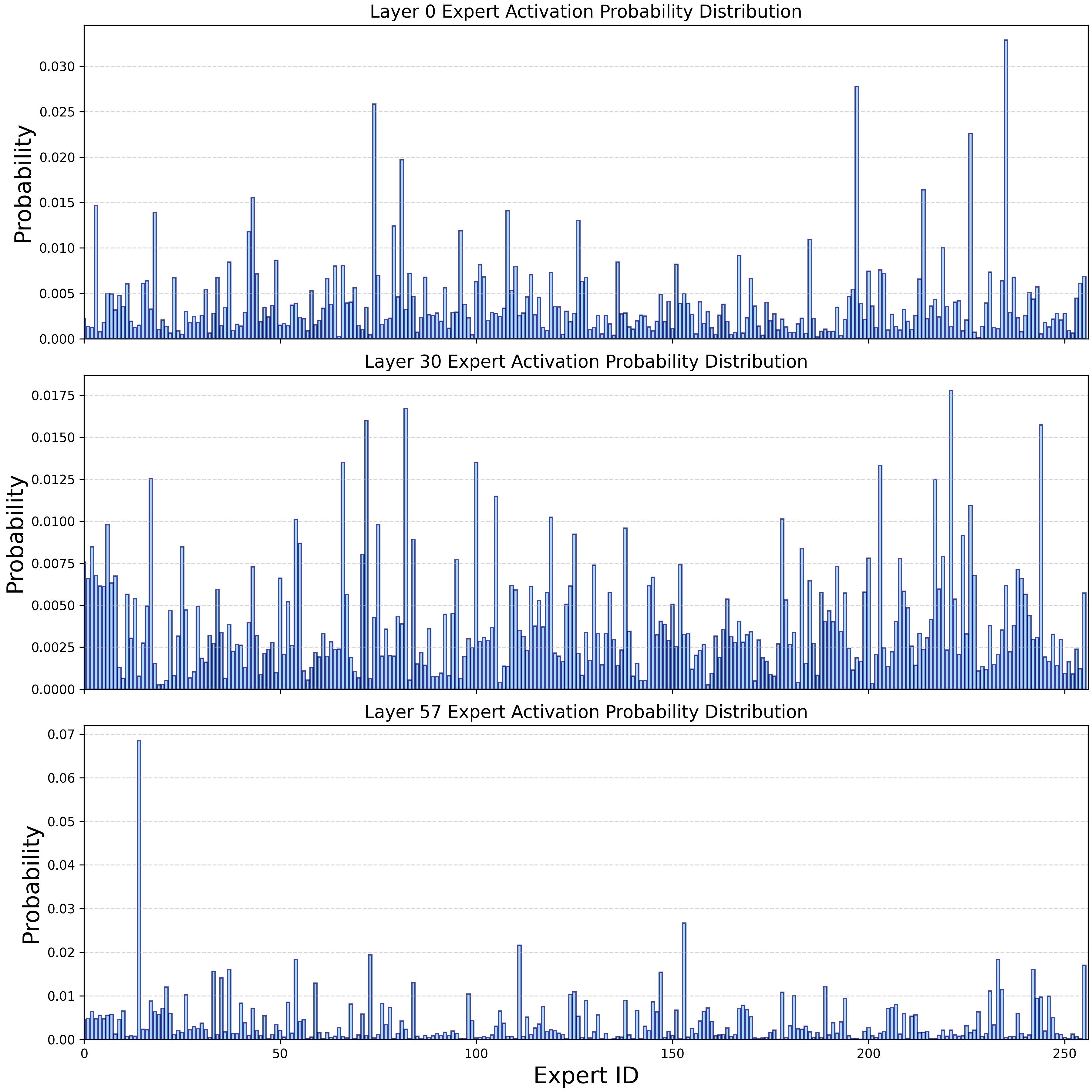}
    \vspace{-5mm}
    \caption{Expert activation distribution of DeepSeek R1  during inference.}
    \label{fig:export_activation}
    \vspace{-3mm}
\end{figure}

\section{Sparse Computing in OmniInfer}
\label{sec:sparse}

\subsection{OmniPlacement: Expert Scheduling for MoE}
\label{subsec:omniplacement}

Mixture-of-Experts inference frequently encounters severe {load imbalance}, characterized by a subset of ``hot'' experts receiving a disproportionate share of activations, as shown in Figure~\ref{fig:export_activation}. This skew creates computational bottlenecks on specific devices, thereby increasing end-to-end latency and throttling system throughput.

To mitigate this, we propose {OmniPlacement}, a dynamic, near real-time expert placement algorithm. OmniPlacement exploits expert redundancy and a lightweight scheduling mechanism to rapidly adapt to fluctuating workload patterns. Our approach is designed to maintain consistently balanced expert utilization, thereby substantially enhancing MoE inference performance.

\subsubsection{Problem Formulation}
We consider an MoE model comprising $L$ layers, with $E$ experts per layer, deployed across a cluster of $R$ devices. The expert placement is represented by a binary tensor $P \in \{0, 1\}^{L \times R \times E}$, where $P_{l, r, e}=1$ indicates that expert $e$ of layer $l$ is hosted on device $r$.

\emph{Constraints.} The placement is subject to two primary constraints. First, to ensure model integrity, every expert must be available on at least one device:
\begin{equation}
    \sum_{r=0}^{R-1} P_{l, r, e} \ge 1, \quad \forall l, e.
    \label{eq:existence_constraint}
\end{equation}
Second, we impose a capacity constraint to manage memory usage. Each device allocates a fixed budget of $s_l$ expert ``slots'' for layer $l$. The parameter $s_l$, termed the redundancy factor, dictates the replication level. This constraint is expressed as:
\begin{equation}
    \sum_{e=0}^{E-1} P_{l, r, e} \le s_l, \quad \forall l, r.
    \label{eq:capacity_constraint}
\end{equation}
A larger $s_l$ increases the redundancy budget ($s_l R - E$ total replicas), allowing ``hot'' experts to be replicated across multiple devices to distribute load, albeit at the cost of higher memory consumption.

\begin{algorithm}[tb]
  \LinesNumbered
  \small
  \caption{Static Expert Placement}
  \label{alg:static_placement}
  \KwIn{Load matrix $D$, number of layers $L$, number of devices $R$, total redundancy budget $M$}
  \KwOut{Optimal static placement $P^*$}
  
  $s \gets \texttt{AllocateBudgetByImbalance}(D, L, M)$\;
  
  \For{$l = 0$ \KwTo $L-1$}{
    $P^{\text{best}}_{l} \gets \text{null}, B_{\text{min}} \gets \infty$\;
    \For{$k = 0$ \KwTo $s_l$}{
      $C \gets \texttt{DetermineReplicas}(D_{l}, k, R)$\;
      $P^k \gets \texttt{GeneratePlacement}(C, D_{l}, R)$\;
      $B_k \gets \texttt{CalculateImbalance}(P^k, D_{l})$\;
      \If{$B_k < B_{\text{min}}$}{
        $P^{\text{best}}_{l} \gets P^k$\;
        $B_{\text{min}} \gets B_k$\;
      }
    }
    $P^*_{l} \gets P^{\text{best}}_{l}$\;
  }
  \Return{$P^*$}\;
\end{algorithm}

\emph{Objective.} Our goal is to minimize load imbalance. To formalize this, we define the expert load matrix $D \in \mathbb{R}^{L \times E}_{\ge 0}$. Each entry $D_{l,e}$ represents the estimated computational load of expert $e$ in layer $l$. This value is derived from historical activation statistics (e.g., via a sliding window average) to capture recent workload trends while filtering out transient noise.
Given the placement $P$ and load matrix $D$, the total load on device $r$ for layer $l$ is the aggregate load of its hosted experts:
\begin{equation}
    R_r(l, P, D) = \sum_{e=0}^{E-1} P_{l,r,e} \cdot D_{l,e}.
    \label{eq:device_load}
\end{equation}
We define the \emph{load imbalance ratio} $B(l, P, D)$ as the ratio of the peak device load to the average device load:
\begin{equation}
    B(l, P, D) = \frac{\max_{i} R_i(l, P, D)}{\frac{1}{R} \sum_{j} R_j(l, P, D)}.
    \label{eq:imbalance_metric}
\end{equation}
A ratio of 1.0 indicates perfect balance. 
Our objective is to find a placement $P$ that minimizes the load imbalance ratio
subject to the existence constraint in Equation~\eqref{eq:existence_constraint} and the capacity constraint in Equation~\eqref{eq:capacity_constraint}.
Moreover, we empirically illustrate the evolution of the load imbalance ratio under real-world workloads in Figure~\ref{fig:imbalance_ratio}.

\subsubsection{OmniPlacement Algorithms}
To achieve the goals, our expert management strategy, OmniPlacement, consists of two core components. 
First, a static placement algorithm (Algorithm~\ref{alg:static_placement}) generates an optimal initial layout by strategically allocating redundant experts to balance the initial load. 
Second, a dynamic scheduler (Algorithm~\ref{alg:dynamic_scheduler}) continuously monitors the system and performs online adjustments to adapt to shifting workloads. 
Both algorithms are presented below to highlight their primary logic.

The OmniPlacement process begins with the Static Expert Placement algorithm, detailed in Algorithm~\ref{alg:static_placement}. 
Its purpose is to compute an optimal initial expert layout based on a given activation distribution. 
The process starts by intelligently distributing the total memory budget; the \texttt{AllocateBudgetByImbalance} function (line 1) analyzes historical activation data $D$ to assign a larger portion of the total redundancy budget $M$ to layers exhibiting higher load imbalance, ensuring resources are prioritized where they are most needed. 
Next, for each layer, the algorithm seeks the best placement within its allocated budget $s_l$. 
It iterates through possible redundancy levels (line 4) and, for each level, first calls \texttt{DetermineReplicas} (line 5). 
This function employs a heap-based greedy strategy to identify the most frequently activated (``hot'') experts and determines how many replicas of each are needed. 
With the replica counts decided, the \texttt{GeneratePlacement} function (line 6) then maps these expert instances to physical devices. 
This is a two-step process: an initial greedy placement assigns experts to the least-loaded devices, followed by a topology-aware remapping that adjusts the layout to minimize inter-device communication costs. 
By evaluating load imbalance across candidate configurations (lines 7–10), the algorithm selects the placement $P^{*}$ that minimizes imbalance, initializing the system in a well-balanced state.

\begin{algorithm}[tb]
  \LinesNumbered
  \small
  \caption{Dynamic Expert Scheduler}
  \label{alg:dynamic_scheduler}
  \KwIn{Initial placement $P_{\text{init}}$, imbalance trigger threshold $B_{\text{trigger}}$, improvement margin $\Delta$, memory budget $M$}
  \KwOut{Continuously updated placement $P_{\text{current}}$}
  
  $P_{\text{current}} \gets P_{\text{init}}$\;
  $D \gets \texttt{InitializeActivationWindow}()$\;
  
  \While{system is running}{
    $D \gets \texttt{UpdateActivationWindow}()$\;
    
    $B_{\text{current}} \gets \texttt{CalculateImbalance}(P_{\text{current}}, D)$\;
    
    \If{$B_{\text{current}} > B_{\text{trigger}}$}{
      $D_{\text{pred}} \gets \texttt{PredictFutureActivations}(D)$\;
      $P_{\text{cand}} \gets \texttt{StaticExpertPlacement}(D_{\text{pred}}, L, R, M)$\;
      $B_{\text{sim}} \gets \texttt{CalculateImbalance}(P_{\text{cand}}, D_{\text{pred}})$\;
      
      \If{$B_{\text{sim}} < B_{\text{current}} - \Delta$}{
        $\texttt{ExecuteMigration}(P_{\text{current}}, P_{\text{cand}})$\;
        $P_{\text{current}} \gets P_{\text{cand}}$\;
      }
    }
    
    Wait for next scheduling interval\;
  }
\end{algorithm}

Once the system is running, the {Dynamic Expert Scheduler} (Algorithm~\ref{alg:dynamic_scheduler}) takes over to perform continuous, online optimization. 
The scheduler monitors real-time expert activations and updates a sliding window of workload data (line 4). 
If the current load imbalance $B_{\text{current}}$ surpasses a trigger threshold $B_{\text{trigger}}$ (line 6), it initiates a re-balancing procedure. 
To make proactive decisions, it first calls \texttt{PredictFutureActivations} (line 7), a function that analyzes recent activation trends to forecast the workload $D_{\text{pred}}$ for the upcoming time interval.
This predictive approach allows the system to adapt to shifting patterns before they become critical bottlenecks. 
Using this predicted workload, a new candidate placement $P_{\text{cand}}$ is generated by invoking the static placement algorithm (line 8). 
To prevent system instability from frequent, minor adjustments, this new placement is only adopted if its simulated imbalance $B_{\text{sim}}$ shows a significant improvement over the current state, exceeding a predefined margin $\Delta$ (line 10). 
If the change is accepted, the \texttt{ExecuteMigration} function is called (line 11) to apply the new layout. 
In essence, this function orchestrates the physical migration of expert weights, ensuring a seamless transition with minimal impact on ongoing inference tasks.

\subsection{OmniAttn: Sparse Attention Acceleration}
\label{subsec:omniattn}
The attention patterns of large language models (LLMs) are often inherently sparse: many attention heads consistently focus on only a small subset of key tokens, while the remaining positions contribute marginally to the output. 
Recent empirical studies, such as {Streaming LLM} \cite{xiao2024efficient} and {Duo Attention} \cite{xiao2024duoattentionefficientlongcontextllm}, reveal that attention is typically concentrated on the earliest tokens (i.e., \emph{attention sink}) and the most recent tokens (i.e., \emph{recent}), whereas intermediate tokens receive near-zero attention scores. 
This observation enables a principled reduction of KV-cache size and attention computation cost without significantly compromising generation quality. Capitalizing on this, we propose OmniAttn, a KV cache compression algorithm that employs a layer-wise pruning pattern and a fast pattern search method based on an evolutionary algorithm.

\subsubsection{Problem Formulation.} 
Given a query matrix $Q \in \mathbb{R}^{K \times d}$, and key and value matrices $K, V \in \mathbb{R}^{M \times d}$, we aim to identify a small subset of token indices $\mathcal{M} \subseteq \{1, 2, \ldots, M\}$ with size $N = |\mathcal{M}| \ll M$.

Let $K_{\mathcal{M}} \in \mathbb{R}^{N \times d}$ and $V_{\mathcal{M}} \in \mathbb{R}^{N \times d}$ be the matrices formed by selecting the rows from $K$ and $V$ corresponding to the indices in $\mathcal{M}$. The goal is to find an $\mathcal{M}$ such that the following approximation holds:
\begin{equation}
    \mathrm{softmax}\left(\frac{Q K_{\mathcal{M}}^\top}{\sqrt{d}}\right) V_{\mathcal{M}}
    \approx 
    \mathrm{softmax}\left(\frac{Q K^\top}{\sqrt{d}}\right) V.
\end{equation}

In {OmniAttn}, $\mathcal{M}$ is constructed by selecting tokens from the \emph{attention sink region} and the \emph{recent region}, i.e.,
\begin{equation}
\mathcal{M} \coloneqq \{1, 2, \ldots, N_{\text{sink}}\} \cup \{M - N_{\text{recent}} + 1, \ldots, M\},
\end{equation}
where $N_{\text{sink}}$ and $N_{\text{recent}}$ are hyperparameters controlling compression granularity. In contrast to methods such as StreamingLLM \cite{xiao2024efficient}, we apply pattern search for layer-wise attention compression as described in the following.

\subsubsection{Layer-wise Compression.} 
While prior sparse attention methods typically apply compression at the {attention head} level, this introduces {imbalanced head effects} under tensor parallelism (TP), where compressed heads must wait for the slowest uncompressed head to complete. 
To address this, OmniAttn performs compression at the {layer level}, treating each Transformer layer as the minimal compression unit. 
This design eliminates synchronization delays under TP and is also compatible with Multi-Latent Attention (MLA)–style architectures\cite{liu2024deepseekv2,ji2025towards}. 
However, applying a single pruning strategy to all heads within a layer presents a greater challenge in maintaining model accuracy. 
To address this, we introduce a fast pattern search algorithm that efficiently and thoroughly explores the configuration space at an inference-only cost.

\subsubsection{Pattern Search.} 
Unlike prior works like DuoAttention~\cite{xiao2024duoattentionefficientlongcontextllm} that rely on end-to-end training to learn pruning patterns -- a process that is computationally and memory intensive due to the need for gradient computation -- OmniAttn adopts a search-based strategy. 
We leverage an evolutionary algorithm to discover high-performance layer-wise compression patterns at an inference-only cost, avoiding any training process.

Specifically, we use a genetic algorithm to minimize inference latency under an accuracy constraint,
Each candidate pattern is represented as a binary vector $p\in\{0,1\}^L$, where each element indicates whether the corresponding layer is compressed. The search proceeds as follows:

\begin{enumerate}
    \item Initialization: A population of candidate patterns is generated randomly.
    \item Evaluation: Each pattern is used to configure the model's KV cache, and the model's accuracy is evaluated on a held-out dataset. This accuracy score serves as the pattern's fitness.
    \item Evolution: The population undergoes selection, crossover (evolution), and mutation over multiple generations, favoring patterns with higher fitness scores.
\end{enumerate}

The search terminates after a fixed number of generations or upon early stopping if a pattern exceeds the target accuracy threshold $\tau$. This process can be formalized as:
\begin{equation}
\min_{p} \ \mathrm{Latency}(p) \quad \text{s.t.} \quad \mathrm{Acc}(p) \geq \tau,
\end{equation}
where $p$ is the layer-wise compression pattern.

\section{Smart Scheduling in OmniInfer}
\label{sec:scheduling}

\subsection{OmniProxy: Global PD Adaptive Scheduler}
\label{subsec:omniproxy}

To make globally optimal scheduling decisions under heterogeneous and dynamic workloads, {OmniProxy} acts as the central scheduling layer that coordinates both the {prefill} and {decode} phases.
It integrates system-level lifecycle management with adaptive scheduling algorithms, enabling fine-grained request reordering, batching, and load distribution while minimizing tail latency.

\subsubsection{System Framework.}
OmniProxy is built on top of the high-performance network service framework \emph{Nginx}, and extends it with three core capabilities:

\emph{Unified Request Lifecycle.} A unified request lifecycle management layer is introduced to coordinate execution across the Prefill and Decode stages. 
It enables flexible state transitions, fine-grained timing control, and cache-aware routing. 
The full lifecycle comprises eight phases: tokenize,  Automatic Prefix Cache (APC) matching, prefill waiting, prefill scheduled, prefill running, decode waiting, decode scheduled, and decode running.

\emph{Real-Time Performance Metrics Collecting.} OmniProxy continuously gathers both request-level signals (e.g., prompt length, prefix-match score, TPOT, TTFT) and instance-level metrics (e.g., queue length, batch execution time, throughput).
These real-time measurements enable predictive and feedback-driven scheduling.

\emph{Deferred Submission and Resorting.} Unlike Nginx, which schedules and dispatches requests immediately upon arrival, OmniProxy defers submission to the backend and performs dynamic resorting. 
By incorporating the predicted upstream batch cycle into the scheduling process, the system ensures that deferral introduces minimal impact on TTFT. 
This postponed submission provides additional flexibility to form more coherent request groups and achieve more balanced load balancing across the prefill and decode stages.

\subsubsection{Omni Adaptive Scheduling (OAS) Algorithm.}

Built upon this system framework, OmniProxy proposes an Omni Adaptive Scheduling (OAS) algorithm that uses both request-level and instance-level signals to achieve globally coordinated scheduling:

\emph{Prefill Side: Cache-Informed Load-Balanced Scheduling.}
OmniProxy integrates with the inference engine’s Automatic Prefix Cache (APC) mechanism~\cite{vllm2025apc}. 
For an incoming request $i$, OmniProxy evaluates each candidate prefill node using the scheduling score,
\begin{equation}
\pi_{P}(i) = \text{Match}_{P}(i) - \alpha \rho_{P},
\end{equation}
where $\text{Match}_{P}(i)$ denotes the prefix match score obtained via radix-tree lookup, and $\rho_{P}$ represents the instantaneous load of the prefill node, i.e., the number of running requests and tokens. 
The coefficient $\alpha$ controls the trade-off between cache locality and load balancing.
This formulation enables OmniProxy to maximize cache reuse while preventing load concentration on any single node.

\emph{Decode Side: Processing-Time–Oriented Scheduling.}
For the decode stage, OmniProxy estimates the effective workload of request $i$ as
\begin{equation}
\ell_i = T_i^{\mathrm{prompt}} + T_i^{\mathrm{max}},
\end{equation}
where $T_i^{\mathrm{prompt}}$ and $T_i^{\mathrm{max}}$ denote prompt length and max tokens if provided. Requests are then scheduled according to a Longest-Processing-Time-First (LPT) policy to minimize execution gaps and improve throughput.

\section{Evaluation}
\label{sec:evaluation}
We conduct a comprehensive experimental evaluation to assess the efficiency, scalability, and accuracy of {OmniInfer} under realistic large-scale inference workloads.
Our experiments are designed to answer three key questions: (i) \emph{How well does OmniInfer scale with different prefill and decode configurations?} (ii) \emph{How much does each acceleration component contribute to end-to-end performance?} and (iii) \emph{Does the use of sparse attention affect model accuracy?}
All experiments are conducted on realistic hardware configurations and widely used LLM models, ensuring the results reflect practical deployment conditions.

\begin{table}[tbp]\footnotesize
% \vspace{-2mm}
% \scriptsize
% \small
\centering
\setlength{\tabcolsep}{3.5pt}
\caption{Scaling experiment results under different xPyD configurations.}
\vspace{-3mm}
\label{tab:scaling_results}
\resizebox{0.45\textwidth}{!}{
\begin{tabular}{lcccc}
\toprule
\textbf{xPyD} & \textbf{Batch Size/die} & \textbf{QPM} (r/s) & \textbf{TTFT (s)} & \textbf{TPOT (ms)} \\
\midrule
4P8-1D32 & 24 & 472 & 2.598 & 38 \\
5P8-1D32 & 30 & 500 & 1.548 & 41 \\
5P8-1D32 & 32 & 520 & 1.983 & 42 \\
6P8-1D32 & 40 & 524 & 1.282 & 48 \\
6P8-1D32 & 44 & 552 & 1.630 & 50 \\
6P8-1D32 & 46 & 549 & 1.537 & 51 \\
6P8-1D32 & 48 & 572 & 2.196 & 52 \\
8P8-1D64 & 24 & 481 & 2.822 & 37 \\
\bottomrule
\end{tabular}}
\end{table}

\begin{table*}[htbp]\footnotesize                                                          
\centering                                                                   
\setlength{\tabcolsep}{4pt}
\caption{End-to-End performance of OmniInfer.}
\vspace{-3mm}
\resizebox{\textwidth}{!}{
\begin{tabular}{lccccccccc}
\toprule
\textbf{Method} & \textbf{QPM}(r/s) & \textbf{TPOT}(ms) & \textbf{p99 TPOT}(ms) & \textbf{TTFT}(s) & \textbf{p99 TTFT}(s) & \textbf{E2E}(s) & \textbf{p99 E2E}(s) & \textbf{OTT}(tok/s) & \textbf{TTT}(tok/s) \\
\midrule
OmniInfer & 616 & 48 & 50 & 2.292 & 4.198 & 59.206 & 148.364 & 60,575 & 225,546 \\
w/o OmniPlacement & 463 & 65 & 68 & 1.342 & 2.419 & 78.025 & 196.226 & 45,372 & 169,083 \\
w/o OmniAttn & 554 & 55 & 57 & 1.237 & 2.174 & 65.791 & 161.141 & 54,533 & 202,628  \\
w/o OmniProxy & 575 & 50 & 61 & 3.683 & 10.711 & 63.451 & 157.843 & 56,969 & 211,279 \\
w/o all & 404 & 75 & 101 & 1.224 & 3.374 & 88.169 & 213.846 & 39,954 & 149,500  \\
\bottomrule
\end{tabular}}
\label{tab:ablation}
\end{table*}

\begin{table}[htbp]
% \footnotesize
\centering
\setlength{\tabcolsep}{3pt}
\caption{Accuracy of OmniAttn on five reasoning benchmarks.}
\vspace{-3mm}
\label{tab:accuracy_results}
\resizebox{0.48\textwidth}{!}{
\begin{tabular}{lccccc}
\toprule
\textbf{Method} & \textbf{AIME24} & \textbf{MMLU-Pro} & \textbf{GSM8K} & \textbf{LongBench-V2} & \textbf{MATH500} \\
\midrule
DS-R1 & 79.80 & 84.00 & 71.50 & 40.80 & 87.20 \\
+ OmniAttn & 80.28 & 84.38 & 69.19 & 41.90 & 87.80 \\
\bottomrule
\end{tabular}
}
\end{table}

\subsection{Experimental Setup}
All experiments are conducted on {Ascend~910C} NPUs using {DeepSeek-R1} models with {INT8 quantization}. 
We evaluate the performance of OmniInfer under different prefill and decode configurations.
For the {prefill (P)} stage, each Ascend~910C node employs \texttt{16}-way tensor parallelism.%
\footnote{Each Ascend~910C device consists of two dies, resulting in 16 dies per node.} 
We use the \texttt{P8} configuration, meaning that one Ascend~910C node hosts a single {DeepSeek-R1} model with TP16.
For the {decode (D) stage, we evaluate two configurations: \texttt{D32} and \texttt{D64}, corresponding to \texttt{64}-way and \texttt{128}-way data parallelism, respectively. 
These settings involve four-node and eight-node Ascend~910C nodes per model. 
The batch size reported in our experiments refers to the \texttt{per-die batch size}. 
The total system-wide batch size is computed as:
$\texttt{system-level batch size} = \texttt{per-die batch size} \times \#\texttt{NPU nodes} \times 8 \times 2$.
For example, under the \texttt{4P8-1D32} configuration with a per-die batch size of 24, the system consists of four prefill nodes, each hosting a single {DeepSeek-R1} model with TP16, and one decode node composed of four Ascend~910C nodes serving a {DeepSeek-R1} model with DP64. 
The total concurrency for this system is 1536.

Performance is benchmarked using a variable-length long-context workload with an average input length of 3.5K tokens and an average output length of 1K tokens.
We measure both throughput and latency metrics, including Queries Per Minute (QPM), Time-To-First-Token (TTFT), Time-Per-Output-Token (TPOT), Output Token Throughput (OTT), and Total Token Throughput (TTT).
Detailed descriptions of the dataset construction, benchmarking methodology, and system configurations are provided in the Appendix~\ref{app:experiment}.

\subsection{Experimental Results}

\subsubsection{Scaling Experiments}
To investigate the impact of system configuration on serving efficiency, we conduct scaling experiments by varying  the ratio of prefill and decode nodes (\texttt{xPyD}) and  the batch size. 
The results are summarized in Table~\ref{tab:scaling_results}.
From the results, several observations can be made.
\begin{itemize}[leftmargin=*]
\item \emph{Batch size scaling.} Larger batch sizes initially improve throughput, but beyond a saturation point TTFT grows disproportionately, indicating the presence of an optimal batching window.

\item \emph{Effect of P/D ratio.} The ratio between prefill and decode nodes strongly affects TTFT: higher decode capacity requires additional prefill nodes to maintain balanced utilization and meet SLOs.

\item \emph{Scaling inefficiency.} Increasing prefill or decode nodes alone does not scale QPM linearly. For example, \texttt{8P8-1D64} provides only marginal gains over \texttt{4P8-1D32} despite doubling resources.

\item \emph{Data-parallel size.} Larger decode-side data parallelism reduces TPOT but exhibits diminishing returns, consistent with the system-wide throughput saturation trend.
\end{itemize}

\subsubsection{End-to-End Performance Results}
We perform the final performance evaluation of our system using the best-performing configuration identified in the scaling experiments, namely \texttt{6P8-1D32}. All proposed components are integrated in this setting, and we further perform an ablation study to quantify the individual contributions of each module.
Table~\ref{tab:ablation} presents the tail-latency, throughput, and efficiency metrics across different ablation variants. Several key observations can be drawn:
\begin{itemize}[leftmargin=*]
\item \emph{Overall improvement.}
The baseline system (w/o all) reaches only 404 QPM, whereas the full OmniInfer configuration achieves 616 QPM—representing a \textbf{52\%} increase in end-to-end throughput.
\item \emph{Impact of sparse-computing modules.}
Removing {OmniPlacement} or {OmniAttn} reduces throughput to 463 and 554 QPM, respectively. These modules address MoE and attention sparsity, and their removal leads to inefficient batching and increased token generation cost. Quantitatively, they contribute approximately {33\%} and {11\%} throughput gains.
\item \emph{TPOT–TTFT trade-off at high throughput.}
Higher throughput increases the prefill-side pressure and raises TTFT. For instance, TTFT grows from $\SI{1.342}{\second}$ (w/o OmniPlacement) and $\SI{1.237}{\second}$ (w/o OmniAttn) to $\SI{3.683}{\second}$ when OmniProxy is disabled.
\item \emph{Role of OmniProxy as the balancing layer.}
With OmniProxy, the system increases throughput from 575 (w/o OmniProxy\footnote{We use the default round-robin load balancer in Nginx when disabling OmniProxy.}) to 616 QPM, while preventing excessive TTFT growth. This highlights OmniProxy's role as a global scheduler that harmonizes prefill and decode and ensures both efficiency and responsiveness.
\end{itemize}

\subsubsection{Accuracy Results}
To evaluate the impact of KV-cache compression on model quality, we assess the accuracy of {OmniAttn} across five representative reasoning benchmarks: AIME24~\cite{AIME2024}, MMLU-Pro~\cite{wang2024mmlu}, GSM8K~\cite{cobbe2021training}, LongBench-V2~\cite{bai2025longbench}, and MATH500~\cite{hendrycks2021measuring}. Our implementation of OmniAttn accelerates inference by retaining only critical tokens (i.e., attention sinks and recent tokens), thereby reducing the size of the KV cache without modifying model weights.
Table~\ref{tab:accuracy_results} summarizes the evaluation results. Across all benchmarks, OmniAttn achieves comparable accuracy relative to the baseline model, demonstrating that KV compression does not lead to significant degradation in reasoning performance. These results indicate that sparse attention patterns can be exploited to improve efficiency while maintaining accuracy, making OmniAttn a practical optimization for large-scale inference.

\section{Conclusion and Future Work}
\label{sec:conclusion}

We present OmniInfer, a modular and hardware-co-designed system for efficient LLM serving based on a globally coordinated, disaggregated architecture.
Through the synergy of OmniPlacement, OmniAttn, and OmniProxy, OmniInfer resolves the prefill–decode tension and achieves state-of-the-art performance on Ascend NPUs.
Extensive evaluation confirms substantial gains in throughput and latency, demonstrating a scalable path toward next-generation serving infrastructure.

While OmniInfer demonstrates significant performance gains, several directions remain for future work. First, extending support to a broader range of LLMs, acceleration techniques, and multimodal models to further generalize the system. Second, integrating OmniInfer into additional inference frameworks beyond vLLM, leveraging its modular design to facilitate seamless deployment and improvement. Third, adapting OmniInfer to other hardware platforms beyond Ascend NPUs, enabling wider applicability and cross-platform acceleration.

\bibliographystyle{ACM-Reference-Format}
\bibliography{sample-base}

\appendix

\section{Implementation Details of OmniPlacement}
\label{app:omniplacement}

\subsection{Near Real-time Scheduling and Dynamic Monitoring Mechanism}
Near real-time expert activation serves as the foundation for dynamic balancing. Based on the Ascend platform, we designed a near real-time scheduling and dynamic monitoring mechanism that collects and analyzes expert activation information while maintaining inference performance. 
\paragraph{Layer-Wise Redundant Deployment}
OmniPlacement adaptively allocates replicas based on the imbalance ratio $B(l, P, D)$. Layers exhibiting high imbalance receive additional replicas for their ``hot'' experts, while stable layers remain compact. This heterogeneous resource distribution directs memory overhead specifically toward bottleneck layers, offering superior robustness against load spikes compared to uniform redundancy schemes.

\paragraph{Near Real-Time Scheduling and Monitoring}
To facilitate dynamic adaptation, OmniPlacement employs an asynchronous monitoring component running on a separate computation stream. This component tracks expert activations in real-time without blocking the main inference process. The load $D_{l,e}$ is computed via a weighted moving average, effectively filtering transient noise. This data drives the imbalance calculations in Equation~(\ref{eq:imbalance_metric}), enabling the scheduler to rapidly converge to near-optimal configurations.

\paragraph{Pipelined Expert Weight and Placement Updates}
We implement a sophisticated pipeline to decouple weight migration from inference. While inference proceeds on the main stream, expert weights are transferred in the background using a dedicated communication stream, leveraging primitives such as Huawei's Collective Communication Library (HCCL) for low-latency transport. The update concludes with an atomic switch to the new configuration once migration is complete. This pipelined approach enables seamless updates, ensuring that migration overhead has a negligible impact on inference latency.

\subsection{Load Imbalance Evaluation}
We further evaluate the effectiveness of OmniPlacement in mitigating expert load imbalance by comparing it against EPLB on a production workload. 
Specifically, we measure the layer-wise load imbalance ratio defined in Equation~(\ref{eq:imbalance_metric}) throughout the inference process.
As shown in Figure~\ref{fig:imbalance_ratio}, the original system exhibits severe load skew, with the maximum imbalance ratio reaching nearly 8, indicating that a small number of hot experts dominate the computation and lead to persistently unbalanced execution. 
In contrast, OmniPlacement consistently maintains a substantially lower imbalance ratio, approaching the ideal value of 1 (dotted line), by dynamically replicating hot experts and redistributing workload across devices.
Moreover, the near real-time monitoring mechanism enables OmniPlacement to rapidly adapt to workload shifts and transient activation bursts, achieving more stable and balanced execution compared to EPLB.
Overall, these results demonstrate that OmniPlacement effectively mitigates expert load imbalance and delivers robust performance improvements under realistic serving conditions.

\begin{figure}[ht]  
    \centering
    \includegraphics[width=1\linewidth]{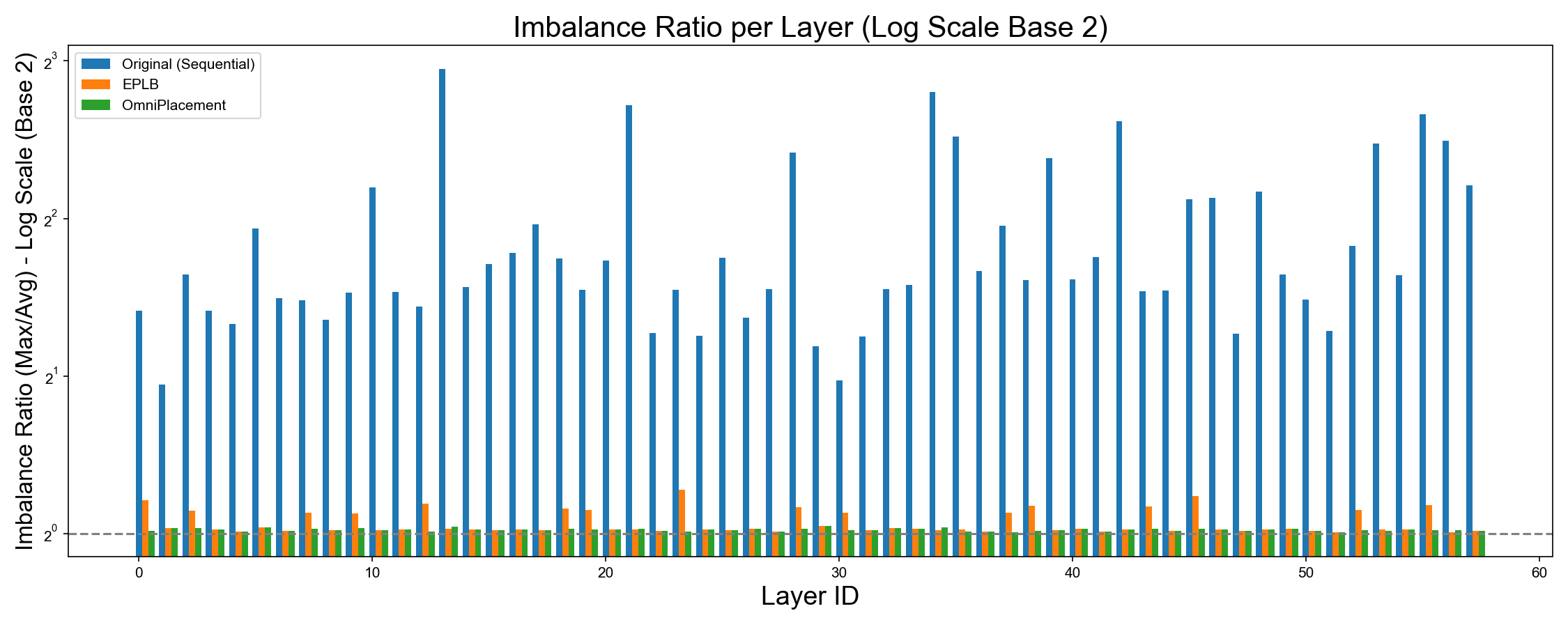}
    \vspace{-5mm}
    \caption{Comparison of layer-wise expert load imbalance ratios.}
    \label{fig:imbalance_ratio}
    \vspace{-3mm}
\end{figure}

\subsection{Configuration of OmniPlacement}
In our deployment, each device hosts at most one additional replicated expert beyond the baseline placement.
When dynamic replication is enabled, the memory overhead scales linearly with the expert size, the number of redundant experts per device, and the MoE layer number, enabling a flexible trade-off between memory consumption and load balancing effectiveness.

\section{Implementation Details of Experiment}
\label{app:experiment}

\paragraph{Datasets.}
We construct a synthetic dataset derived from large-scale online novel corpora to emulate real-world long-context generation scenarios.
The dataset features variable-length input and output sequences, with an average input length of approximately 3{,}500 tokens and an average output length of about 1{,}000 tokens.
The combined sequence length (input plus output) is capped below 16K tokens.
Both input and output length distributions exhibit a pronounced long-tail pattern, reflecting the heterogeneous nature of real LLM serving workloads.

\paragraph{Benchmarking.} 
To impose sustained and controllable system load, we design a customized benchmarking script that maintains a fixed concurrency level throughout the evaluation. 
Specifically, if $n$ requests are completed during a given time interval, the benchmarking system immediately injects $n$ new requests to keep the number of in-flight requests constant. 
This ensures that the serving system operates under steady-state pressure, enabling a fair comparison of scheduling strategies and resource utilization efficiency.
We measure throughput and latency metrics, including Queries Per Minute (QPM), Time-To-First-Token (TTFT), and Time-Per-Output-Token (TPOT), Output Token Throughput (OTT), and Total Token Throughput (TTT). 

\end{document}